\documentclass[prd,twocolumn,showpacs,preprintnumbers,floats]{revtex4}\def\@cite#1#2{\textsuperscript{[{#1\if@tempswa , #2\fi}]}}

\usepackage{graphicx}
\usepackage{amsmath}
\usepackage{amsfonts}
\usepackage{amssymb}
\usepackage{color}
\usepackage{subfigure}
\usepackage{epsfig}
\usepackage{morefloats}
\usepackage{multirow}
\usepackage{graphicx,booktabs}
\usepackage{mathrsfs}
\usepackage{txfonts}
\usepackage{indentfirst}
\usepackage{graphicx,booktabs}

\usepackage{longtable,lscape}

\newcommand{\vsig}{\mbox{\boldmath$\sigma$\unboldmath}}

\newcommand{\vlab}{\mbox{\boldmath$\lambda$\unboldmath}}
\newcommand{\vxi}{\mbox{\boldmath$\xi$\unboldmath}}

\begin{document}

\title{All-heavy tetraquarks }
\author{Ming-Sheng Liu$^{1,4}$~\footnote {E-mail: liumingsheng0001@126.com}, Qi-Fang L\"{u}$^{1,4}$~\footnote {E-mail: lvqifang@hunnu.edu.cn},
Xian-Hui Zhong$^{1,4}$~\footnote {E-mail: zhongxh@hunnu.edu.cn}, Qiang Zhao$^{2,3,4}$~\footnote {E-mail: zhaoq@ihep.ac.cn}}

\affiliation{ 1) Department
of Physics, Hunan Normal University, and Key Laboratory of
Low-Dimensional Quantum Structures and Quantum Control of Ministry
of Education, Changsha 410081, China }

\affiliation{ 2) Institute of High Energy Physics and Theoretical Physics Center for Science Facilities,
Chinese Academy of Sciences, Beijing 100049, China}

\affiliation{ 3)  School of Physical Sciences, University of Chinese Academy of Sciences, Beijing 100049, China}

\affiliation{ 4)  Synergetic Innovation Center for Quantum Effects and Applications (SICQEA),
Hunan Normal University, Changsha 410081, China}

\begin{abstract}

In this work, we study the mass spectra of the all-heavy tetraquark systems, i.e, $cc\bar{c}\bar{c}$, $bb\bar{b}\bar{b}$, $bb\bar{c}\bar{c}/cc\bar{b}\bar{b}$, $bc\bar{c}\bar{c}/cc\bar{b}\bar{c}$, $bc\bar{b}\bar{b}/bb\bar{b}\bar{c}$, and $bc\bar{b}\bar{c}$, within a potential model by including the linear confining potential, Coulomb potential, and spin-spin interactions. It shows that the linear confining potential has important contributions to the masses and is crucial for our understanding of the mass spectra of the all-heavy tetraquark systems. For the all-heavy tetraquarks $Q_1Q_2\bar{Q}_3\bar{Q}_4$, our explicit calculations suggest that no bound states can be formed below the thresholds of any meson pairs $(Q_1\bar{Q}_3)$-$(Q_2\bar{Q}_4)$ or $(Q_1\bar{Q}_4)$-$(Q_2\bar{Q}_3)$. Thus, we do not expect narrow all-heavy tetraquark states to be existing in experiments.
\end{abstract}

\maketitle

\section{Introduction}{\label{introduction}}

Experimental searches for and theoretical studies of exotic hadrons beyond the conventional quark model are an important test of nonperturbative properties of the strong interaction theory QCD. Since the discovery of quark model~\cite{GellMann:1964nj} and QCD, the progresses on the experimental tools have brought to us a lot of novel phenomena in hadron physics. In particular, during the past 15 years, there have been a sizeable number of candidates for QCD exotics~\cite{Patrignani:2016xqp,Olsen:2017bmm,Lebed:2016hpi,Chen:2016qju,Ali:2017jda,Esposito:2016noz,Guo:2017jvc}.
Interestingly, but also puzzlingly, it shows that the number of exotic candidates is far less than what we have expected for the hadron spectroscopy, where the internal effective degrees of freedom of a hadron may contain quarks and gluons beyond the conventional quark model prescription. Strong evidences for such exotic hadrons include some of those recently observed $XYZ$ states, e.g, $X(3872)$, $Z_c(3900)$, $Z_c(4020)$, $Z_b(10610)$, and $Z_b(10650)$~\cite{Patrignani:2016xqp}. In particular, these charged quarkoniumlike states, $Z_c$ and $Z_b$, contain not only the hidden heavy flavor $c\bar{c}$ or $b\bar{b}$, but also charged light flavors of $u\bar{d}$ or $d\bar{u}$. Since at least four constituent quarks are confined inside these $Z_c$ or $Z_b$ states, it makes them the best candidates for QCD exotic hadrons.

Recently, the tetraquarks of all-heavy systems, such as $cc\bar{c}\bar{c}$ and $bb\bar{b}\bar{b}$,
have received considerable attention with the development of experiments.
If there are stable tetraquark $cc\bar{c}\bar{c}$ and/or $bb\bar{b}\bar{b}$ states, they are most likely to be observed
at LHC~\cite{Eichten:2017ual}. In fact, a search for the tetraquark $bb\bar{b}\bar{b}$ states is being carried out by the LHCb Collaboration although
no confirmed information has been observed~\cite{Aaij:2018zrb}.
Other study interests for physicists arise from the special aspects of the all-heavy tetraquark systems~\cite{Chen:2016jxd}.
They may favor to form genuine tetraquark configurations
rather than loosely bound hadronic molecules, since the light mesons cannot be exchanged between two heavy mesons.
Furthermore, it will be very easy to distinguish the all-heavy tetraquark states from the states which have been observed because
their masses should be far away from the mass regions of the observed states. Thus, besides some previous works on the all-heavy tetraquark states~\cite{Ader:1981db,Iwasaki:1975pv,Zouzou:1986qh,Heller:1985cb,Lloyd:2003yc,Barnea:2006sd}, many new studies have been carried out in recent years~\cite{Wang:2017jtz,Karliner:2016zzc,Berezhnoy:2011xn,Bai:2016int,Anwar:2017toa,Esposito:2018cwh,Chen:2016jxd,
Wu:2016vtq,Hughes:2017xie,Richard:2018yrm,Debastiani:2017msn,Wang:2018poa,Richard:2017vry,Vijande:2009kj}, although some of the conclusions are quite different from each other.
In some works, it is predicted that there exist stable bound tetraquark $cc\bar{c}\bar{c}$ states and/or bound tetraquark $bb\bar{b}\bar{b}$ states
with relatively smaller masses below the thresholds of heavy charmonium pairs~\cite{Wang:2017jtz,Karliner:2016zzc,Berezhnoy:2011xn,Bai:2016int,Anwar:2017toa,Esposito:2018cwh,Chen:2016jxd,Debastiani:2017msn,Wang:2018poa}. Thus, their decays into heavy quarkonium pairs through quark rearrangements will be hindered. In contrast, in some other works it is predicted that there should be no stable bound tetraquark $cc\bar{c}\bar{c}$ and $bb\bar{b}\bar{b}$ states~\cite{Wu:2016vtq,Lloyd:2003yc,Ader:1981db,Hughes:2017xie,Richard:2018yrm} because the predicted masses are large enough for them to decay into heavy quarkonium pairs. To some extent, a better understanding of the possible mass locations is not only crucial for understanding their underlying dynamics, but also useful for experimental searches for their existence.

In this work, we systemically study the mass spectra of the all-heavy tetraquark $Q_1Q_2\bar{Q}_3\bar{Q}_4$ systems
with a potential model widely used in the literature~\cite{Godfrey:2004ya,Eichten:1994gt,Godfrey:1985xj,Swanson:2005,Godfrey:2015dia,Eichten:1978tg,Gupta:1984jb,Kwong:1988ae,Chao:2009,Li:2009zu,
Segovia:2016xqb,Wei-Zhao:2013sta,Lakhina:2006fy,Lu:2016bbk,Li:2010vx,Deng:2016stx,Deng:2016ktl,Song:2015nia,Song:2015fha}. Our purpose is to understand two key issues based on the knowledge collected in the study of heavy quarkonium spectrum. The first one is what a quark potential model can tell about the all-heavy tetraquark system. The second one is where the masses of the ground states could be located if the all-heavy tetraquark states do exist.

At this moment, we do not consider any orbital or radial excitations of the all-heavy tetraquarks. Instead, we would like to address where and how the all-heavy tetraquarks would manifest themselves in their lowest states. For a spectrum of multiquark states, a correct identification of the ground state should be the first step towards a better understanding of the multiquark dynamics in the nonperturbative regime.

The potentials between the quarks, such as the linear confining potential, color Coulomb potential, and spin-spin interactions, are adopted the standard forms of the potential models. The model parameters are determined by fitting the mass spectra of charmonium, bottomonium, and $B_c$ meson.
In our calculations, we find both the confining potential and color Coulomb potential are very crucial for understanding the masses of the all-heavy tetraquarks. The linear confining potential as well as the kinetic energy contributes a quite large positive mass term to the all-heavy tetraquarks $Q_1Q_2\bar{Q}_3\bar{Q}_4$, which leads to a large mass
far above the threshold of the meson pair $Q_1\bar{Q}_3$-$Q_2\bar{Q}_4$ or $Q_1\bar{Q}_4$-$Q_2\bar{Q}_3$,
although the color Coulomb potential contributes a very large negative mass term. As a consequence, we find no bound
all-heavy tetraquarks $Q_1Q_2\bar{Q}_3\bar{Q}_4$ below the threshold of any meson pairs $Q_1\bar{Q}_3$-$Q_2\bar{Q}_4$ or $Q_1\bar{Q}_4$-$Q_2\bar{Q}_3$.

The paper is organized as follows: a brief introduction to the framework is given in Sec.~\ref{Fram}. In Sec.~\ref{Numerical results and discussions}, the numerical results and discussions are presented. A short summary is given in Sec.~\ref{Summary}.

\section{Framework} \label{Fram}

\subsection{Quark model classification} \label{Config}
In the charm and bottom quark sector, there are nine different fully-heavy $Q_1Q_2\bar{Q}_3\bar{Q}_4$ systems: $cc\bar{c}\bar{c}$, $bb\bar{b}\bar{b}$, $bc\bar{c}\bar{c}$, $bc\bar{b}\bar{b}$, $bb\bar{c}\bar{c}$, $bc\bar{b}\bar{c}$, $cc\bar{b}\bar{c}$, $bb\bar{b}\bar{c}$ and $cc\bar{b}\bar{b}$. Note that $cc\bar{b}\bar{c}$, $bb\bar{b}\bar{c}$, and $cc\bar{b}\bar{b}$ are the antiparticles of $bc\bar{c}\bar{c}$, $bc\bar{b}\bar{b}$, and $bb\bar{c}\bar{c}$, respectively.
Thus, we need only consider six systems,  $cc\bar{c}\bar{c}$, $bb\bar{b}\bar{b}$, $bc\bar{c}\bar{c}$, $bc\bar{b}\bar{b}$, $bb\bar{c}\bar{c}$, and $bc\bar{b}\bar{c}$, in our calculations.

To calculate the spectroscopy of a $Q_1Q_2\bar{Q}_3\bar{Q}_4$ system, first we construct the configurations in the space of flavor $\otimes$color$\otimes$spin. Considering the Pauli principle and color confinement for the four-quark system $Q_1Q_2\bar{Q}_3\bar{Q}_4$, we have 12
configurations as follows:
\begin{flalign}
~|1\rangle=|[Q_1Q_2]^6_1[\bar{Q_3}\bar{Q_4}]^{\bar{6}}_1\rangle^0_0,~~~~|2\rangle=|\{Q_1Q_2\}^6_0\{\bar{Q_3}\bar{Q_4}\}^{\bar{6}}_0\rangle^0_0,
\nonumber\\
~|3\rangle=|\{Q_1Q_2\}^{\bar{3}}_1\{\bar{Q_3}\bar{Q_4}\}^3_1\rangle^0_0,~~~~|4\rangle=|[Q_1Q_2]^{\bar{3}}_0[\bar{Q_3}\bar{Q_4}]^3_0\rangle^0_0,
\nonumber\\
~|5\rangle=|[Q_1Q_2]^6_1[\bar{Q_3}\bar{Q_4}]^{\bar{6}}_1\rangle^0_1,~~~~|6\rangle=|[Q_1Q_2]^6_1\{\bar{Q_3}\bar{Q_4}\}^{\bar{6}}_0\rangle^0_1,
\nonumber\\
~|7\rangle=|\{Q_1Q_2\}^6_0[\bar{Q_3}\bar{Q_4}]^{\bar{6}}_1\rangle^0_1,~~~~|8\rangle=|\{Q_1Q_2\}^{\bar{3}}_1\{\bar{Q_3}\bar{Q_4}\}^3_1\rangle^0_1,
\nonumber\\
~|9\rangle=|\{Q_1Q_2\}^{\bar{3}}_1[\bar{Q_3}\bar{Q_4}]^3_0\rangle^0_1,~~~|10\rangle=|[Q_1Q_2]^{\bar{3}}_0\{\bar{Q_3}\bar{Q_4}\}^3_1\rangle^0_1,
\nonumber\\
|11\rangle=|[Q_1Q_2]^6_1[\bar{Q_3}\bar{Q_4}]^{\bar{6}}_1\rangle^0_2,~~~|12\rangle=|\{Q_1Q_2\}^{\bar{3}}_1\{\bar{Q_3}\bar{Q_4}\}^3_1\rangle^0_2,
\nonumber
\end{flalign}
where $\{~\}$ and $[~]$ denote the symmetric and antisymmetric flavor wave functions of the two quarks (antiquarks) subsystems, respectively. The subscripts and superscripts are the spin quantum numbers and representations of the color SU(3) group, respectively. A symmetric spatial wave function is implied for the ground states under investigation.

\begin{table*}[htp]
\begin{center}
\caption{\label{configurations} Configurations of all-heavy tetraquarks.}
\begin{tabular}{cccccccccccc}\hline\hline
~& System              ~~~~~~~& $J^{P(C)}$  ~~~~~~~& \multicolumn{3}{c}{ \underline{~~~~~~~~~~~~~~~~~~~~~~~~~~~~~~~~~~~~~~~~~~~Configuration~~~~~~~~~~~~~~~~~~~~~~~~~~~~~~~~~~~~~~~~~~~~~~} }                        ~&\\
\hline
~& $cc\bar{c}\bar{c}$  ~~~~~~~& $0^{++}$             ~~~~~~~& $|\{cc\}^6_0\{\bar{c}\bar{c}\}^{\bar{6}}_0\rangle^0_0$   ~& $|\{cc\}^{\bar{3}}_1\{\bar{c}\bar{c}\}^3_1\rangle^0_0$
                                                                ~&    $\cdot\cdot\cdot$                                                      ~&\\
~&                     ~~~~~~~& $1^{+-}$             ~~~~~~~& $|\{cc\}^{\bar{3}}_1\{\bar{c}\bar{c}\}^3_1\rangle^0_1$   ~&$\cdot\cdot\cdot$
                                                                ~&         $\cdot\cdot\cdot$                        ~&\\
~&                     ~~~~~~~& $2^{++}$             ~~~~~~~& $|\{cc\}^{\bar{3}}_1\{\bar{c}\bar{c}\}^3_1\rangle^0_2$   ~& $\cdot\cdot\cdot$                     ~&                                   $\cdot\cdot\cdot$                        ~&\\
\hline
~& $bb\bar{b}\bar{b}$  ~~~~~~~& $0^{++}$             ~~~~~~~& $|\{bb\}^6_0\{\bar{b}\bar{b}\}^{\bar{6}}_0\rangle^0_0$   ~& $|\{bb\}^{\bar{3}}_1\{\bar{b}\bar{b}\}^3_1\rangle^0_0$
                                  ~&        $\cdot\cdot\cdot$                           ~&\\
~&                     ~~~~~~~& $1^{+-}$             ~~~~~~~& $|\{bb\}^{\bar{3}}_1\{\bar{b}\bar{b}\}^3_1\rangle^0_1$   ~&
                   $\cdot\cdot\cdot$             ~&     $\cdot\cdot\cdot$                             ~&\\
~&                     ~~~~~~~& $2^{++}$             ~~~~~~~& $|\{bb\}^{\bar{3}}_1\{\bar{b}\bar{b}\}^3_1\rangle^0_2$   ~&
                                        $\cdot\cdot\cdot$       ~&       $\cdot\cdot\cdot$                                       ~&\\
\hline
~& $bb\bar{c}\bar{c}$  ~~~~~~~& $0^{+}$                ~~~~~~~& $|\{bb\}^6_0\{\bar{c}\bar{c}\}^{\bar{6}}_0\rangle^0_0$   ~& $|\{bb\}^{\bar{3}}_1\{\bar{c}\bar{c}\}^3_1\rangle^0_0$
                                                ~&           $\cdot\cdot\cdot$                 ~&\\
~&                     ~~~~~~~& $1^{+}$                ~~~~~~~& $|\{bb\}^{\bar{3}}_1\{\bar{c}\bar{c}\}^3_1\rangle^0_1$   ~&
                          $\cdot\cdot\cdot$                     ~&      $\cdot\cdot\cdot$                             ~&\\
~&                     ~~~~~~~& $2^{+}$                ~~~~~~~& $|\{bb\}^{\bar{3}}_1\{\bar{c}\bar{c}\}^3_1\rangle^0_2$   ~&
       $\cdot\cdot\cdot$      & $\cdot\cdot\cdot$                    ~&       &                    ~&\\
\hline
~& $bc\bar{c}\bar{c}$  ~~~~~~~& $0^{+}$                ~~~~~~~& $|(bc)^6_0\{\bar{c}\bar{c}\}^{\bar{6}}_0\rangle^0_0$      ~& $|(bc)^{\bar{3}}_1\{\bar{c}\bar{c}\}^3_1\rangle^0_0$
                                             ~&            $\cdot\cdot\cdot$                         ~&\\
~&                     ~~~~~~~& $1^{+}$                ~~~~~~~& $|(bc)^6_1\{\bar{c}\bar{c}\}^{\bar{6}}_0\rangle^0_1$      ~& $|(bc)^{\bar{3}}_1\{\bar{c}\bar{c}\}^3_1\rangle^0_1$
                                                                ~& $|(bc)^{\bar{3}}_0\{\bar{c}\bar{c}\}^3_1\rangle^0_1$      ~&\\
~&                     ~~~~~~~& $2^{+}$                ~~~~~~~& $|(bc)^{\bar{3}}_1\{\bar{c}\bar{c}\}^3_1\rangle^0_2$      ~&
                    $\cdot\cdot\cdot$                           ~&       $\cdot\cdot\cdot$                                 ~&\\
\hline
~& $bc\bar{b}\bar{b}$  ~~~~~~~& $0^{+}$                ~~~~~~~& $|(bc)^6_0\{\bar{b}\bar{b}\}^{\bar{6}}_0\rangle^0_0$      ~& $|(bc)^{\bar{3}}_1\{\bar{b}\bar{b}\}^3_1\rangle^0_0$
                                                                ~&            $\cdot\cdot\cdot$                          ~&\\
~&                     ~~~~~~~& $1^{+}$                ~~~~~~~& $|(bc)^6_1\{\bar{b}\bar{b}\}^{\bar{6}}_0\rangle^0_1$      ~& $|(bc)^{\bar{3}}_1\{\bar{b}\bar{b}\}^3_1\rangle^0_1$
                                                                ~& $|(bc)^{\bar{3}}_0\{\bar{b}\bar{b}\}^3_1\rangle^0_1$      ~&\\
~&                     ~~~~~~~& $2^{+}$                ~~~~~~~& $|(bc)^{\bar{3}}_1\{\bar{b}\bar{b}\}^3_1\rangle^0_2$      ~&
                           $\cdot\cdot\cdot$                     ~&         $\cdot\cdot\cdot$                        ~&\\
\hline
~& $bc\bar{b}\bar{c}$  ~~~~~~~& $0^{++}$             ~~~~~~~& $|(bc)^6_1(\bar{b}\bar{c})^{\bar{6}}_1\rangle^0_0$
                                                                ~& $|(bc)^6_0(\bar{b}\bar{c})^{\bar{6}}_0\rangle^0_0$& $\cdot\cdot\cdot$\\
                                                                ~&
                                                                ~&
                                                                ~& $|(bc)^{\bar{3}}_1(\bar{b}\bar{c})^3_1\rangle^0_0$
                                                                ~& $|(bc)^{\bar{3}}_0(\bar{b}\bar{c})^3_0\rangle^0_0$ & $\cdot\cdot\cdot$  \\
~&                     ~~~~~~~& $1^{+-}$             ~~~~~~~& $|(bc)^6_1(\bar{b}\bar{c})^{\bar{6}}_1\rangle^0_1$
                                                                ~& $\frac{1}{\sqrt{2}}|(bc)^6_1(\bar{b}\bar{c})^{\bar{6}}_0\rangle^0_1-|(bc)^6_0(\bar{b}\bar{c})^{\bar{6}}_1\rangle^0_1$& $\cdot\cdot\cdot$\\
                                                                ~&
                                                                ~&
                                                                ~& $|(bc)^{\bar{3}}_1(\bar{b}\bar{c})^3_1\rangle^0_1$
                                                                ~& $\frac{1}{\sqrt{2}}|(bc)^{\bar{3}}_1(\bar{b}\bar{c})^3_0\rangle^0_1-|(bc)^{\bar{3}}_0(\bar{b}\bar{c})^3_1\rangle^0_1$& $\cdot\cdot\cdot$\\
~&                     ~~~~~~~& $1^{++}$             ~~~~~~~& $\frac{1}{\sqrt{2}}|(bc)^6_1(\bar{b}\bar{c})^{\bar{6}}_0\rangle^0_1+|(bc)^6_0(\bar{b}\bar{c})^{\bar{6}}_1\rangle^0_1$
                                                                ~& $\frac{1}{\sqrt{2}}|(bc)^{\bar{3}}_1(\bar{b}\bar{c})^3_0\rangle^0_1+|(bc)^{\bar{3}}_0(\bar{b}\bar{c})^3_1\rangle^0_1$
                                                                ~&
                                       $\cdot\cdot\cdot$  ~&\\
~&                     ~~~~~~~& $2^{++}$             ~~~~~~~& $|(bc)^6_1(\bar{b}\bar{c})^{\bar{6}}_1\rangle^0_2$
                                                                ~& $|(bc)^{\bar{3}}_1(\bar{b}\bar{c})^3_1\rangle^0_2$
                                                                ~&
                                         $\cdot\cdot\cdot$   ~&\\
\hline\hline
\end{tabular}
\end{center}
\end{table*}

It should be emphasized that for the $bc\bar{b}\bar{c}$ systems the $J=1$ states can have both $C=\pm1$, which can be constructed by the linear combinations of $|6\rangle$, $|7\rangle$, $|9\rangle$ and $|10\rangle$,
\begin{flalign}
|6'\rangle=\frac{1}{\sqrt{2}}(|(bc)^6_1(\bar{b}\bar{c})^{\bar{6}}_0\rangle^0_1-|(bc)^6_0(\bar{b}\bar{c})^{\bar{6}}_1\rangle^0_1),\\
|7'\rangle=\frac{1}{\sqrt{2}}(|(bc)^6_1(\bar{b}\bar{c})^{\bar{6}}_0\rangle^0_1+|(bc)^6_0(\bar{b}\bar{c})^{\bar{6}}_1\rangle^0_1),\\
|9'\rangle=\frac{1}{\sqrt{2}}(|(bc)^{\bar{3}}_1(\bar{b}\bar{c})^3_0\rangle^0_1-|(bc)^{\bar{3}}_0(\bar{b}\bar{c})^3_1\rangle^0_1),\\
|10'\rangle=\frac{1}{\sqrt{2}}(|(bc)^{\bar{3}}_1(\bar{b}\bar{c})^3_0\rangle^0_1+|(bc)^{\bar{3}}_0(\bar{b}\bar{c})^3_1\rangle^0_1) \ ,
\end{flalign}
where configurations $|6'\rangle$ and $|9'\rangle$ have $C=-1$, and $|7'\rangle$ and $|10'\rangle$ have $C=+1$. Since the permutation symmetries are lost for
the bc and $\bar{b}\bar{c}$ subsystems, in this work, we use $(~)$ denote no permutation symmetries for these quark pair subsystems.

In Table~\ref{configurations}, all possible configurations and corresponding quantum numbers for the $cc\bar{c}\bar{c}$, $bb\bar{b}\bar{b}$, $bb\bar{c}\bar{c}$, $bc\bar{c}\bar{c}$, $bc\bar{b}\bar{b}$ and $bc\bar{b}\bar{c}$ systems are listed.

\begin{table}[htp]
\begin{center}
\caption{\label{parameters} Quark model parameters used in this work.}
\begin{tabular}{cccccccccccc}\hline\hline
~~~~~~&  $m_c$~(GeV)           &~~~~~~~~~~~~~~~~~~~~~~~~~~~~~~~~~~~~~~1.483~~~~~~~~~~~~~\\
~~~~~~&  $m_b$~(GeV)           &~~~~~~~~~~~~~~~~~~~~~~~~~~~~~~~~~~~~~~4.852~~~~~~~~~~~~~\\
~~~~~~&  ${\alpha_{cc}}$        &~~~~~~~~~~~~~~~~~~~~~~~~~~~~~~~~~~~~~~0.5461~~~~~~~~~~~~~\\
~~~~~~&  ${\alpha_{bb}}$        &~~~~~~~~~~~~~~~~~~~~~~~~~~~~~~~~~~~~~~0.4311~~~~~~~~~~~~~\\
~~~~~~&  ${\alpha_{bc}}$        &~~~~~~~~~~~~~~~~~~~~~~~~~~~~~~~~~~~~~~0.5021~~~~~~~~~~~~~\\
~~~~~~&  ${\sigma_{cc}}$~(GeV) &~~~~~~~~~~~~~~~~~~~~~~~~~~~~~~~~~~~~~~1.1384~~~~~~~~~~~~~\\
~~~~~~&  ${\sigma_{bb}}$~(GeV) &~~~~~~~~~~~~~~~~~~~~~~~~~~~~~~~~~~~~~~2.3200 ~~~~~~~~~~~~~\\
~~~~~~&  ${\sigma_{bc}}$~(GeV) &~~~~~~~~~~~~~~~~~~~~~~~~~~~~~~~~~~~~~~1.3000 ~~~~~~~~~~~~~\\
~~~~~~&  ${b}$ ~(GeV $^2$)    &~~~~~~~~~~~~~~~~~~~~~~~~~~~~~~~~~~~~~~0.1425~~~~~~~~~~~~~\\
\hline\hline
\end{tabular}
\end{center}
\end{table}

\begin{table*}[htp]
\begin{center}
\caption{\label{meson mass} The masses (MeV) of bottomonium mesons. Experimental date are taken from PDG~\cite{Patrignani:2016xqp}.}{
\begin{tabular}{cc|ccccccccccccccc}\hline\hline
~& Meson    ~& $\Upsilon$  ~& $\eta_b$          ~& $\Upsilon(2S)$       ~& $\eta_b(2S)$        ~& $h_b(1P)$    ~& $\chi_{b0}(1P)$    ~& $\chi_{b1}(1P)$    ~& $\chi_{b2}(1P)$   \\ \hline
~& Ours    ~& 9460        ~& 9390              ~& 10024                ~& 10005                ~& 9941         ~& 9859               ~& 9933               ~& 9957\\
~& Expt.      ~& 9460        ~& 9399              ~& 10023                ~& 9999                ~& 9899         ~& 9859               ~& 9893               ~& 9912\\
\hline\hline
\end{tabular}}
\end{center}
\end{table*}

\subsection{Hamiltonian for the multiquark system}{\label{Model Hamiltonian}}

The following nonrelativistic Hamiltonian is adopted for the calculation of the masses of the all-heavy $Q_1Q_2\bar{Q}_3\bar{Q}_4$ system:
\begin{equation}\label{Hamiltonian}
H=(\sum_{i=1}^4 m_i+T_i)-T_G+\sum_{i<j}V_{ij}(r_{ij}),
\end{equation}
where $m_i$ and $T_i$ stand for the constituent quark mass and kinetic energy of the $i$th quark, respectively; $T_G$ stands for the center-of-mass (c.m.) kinetic energy of the $Q_1Q_2\bar{Q}_3\bar{Q}_4$ system; $r_{ij}\equiv|\mathbf{r}_i-\mathbf{r}_j|$ is the distance between the $i$th quark and  $j$th quark; and $V_{ij}(r_{ij})$ stands for the effective potential between the $i$th and  $j$th quark.
In this work, we adopt a widely used potential form for $V_{ij}(r_{ij})$~\cite{Eichten:1978tg,Godfrey:1985xj,Swanson:2005,Godfrey:2015dia,Godfrey:2004ya,Lakhina:2006fy,Lu:2016bbk,Li:2010vx,Deng:2016stx,Deng:2016ktl}, i.e,
\begin{equation}\label{vij}
V_{ij}(r_{ij})=V_{ij}^{OGE}(r_{ij})+V_{ij}^{Conf}(r_{ij}) \ ,
\end{equation}
where $V^{OGE}_{ij}$ stands for the one-gluon-exchange (OGE) potential which describes the short-range quark-quark interactions, while $V^{Conf}_{ij}(r_{ij})$ stands for the confinement potential which describes the long-range interaction behaviors. The form of $V^{OGE}_{ij}$ is given by
\begin{equation}\label{voge}
V^{OGE}_{ij}=\frac{\alpha_{ij}}{4}({\vlab}_i\cdot{\vlab}_j)\left\{\frac{1}{r_{ij}}-\frac{\pi}{2}\cdot\frac{\sigma^3_{ij}e^{-\sigma^2_{ij}r_{ij}^2}}{\pi^{3/2}}\cdot\frac{4}{3m_im_j}({\vsig}_i\cdot{\vsig}_j)\right\},
\end{equation}
where $\vsig_i$ are the Pauli matrices, and $\alpha_{ij}$ stands for the strong coupling strength between two quarks.
If the interaction occurs between two quarks or antiquarks, the $\vlab_i\cdot\vlab_j$ operator appearing in Eq.~(\ref{voge}) is defined as
$\vlab_i\cdot\vlab_j\equiv\sum_{a=1}^8\lambda_i^a\lambda_j^a$, while if the interaction occurs between a quark and antiquark, the $\vlab_i\cdot\vlab_j$ operator is defined as $\vlab_i\cdot\vlab_j\equiv\sum_{a=1}^8-\lambda_i^a\lambda_j^{a*}$, where $\lambda^{a*}$ is the complex conjugate of the Gell-Mann matrix $\lambda^a$. The OGE potential $V^{OGE}_{ij}$ is composed of the Coulomb type potential $V^{OGE}_{coul}\propto (\vlab_i\cdot\vlab_j)(1/r_{ij})$ which provides the short-range interaction, and the color-magnetic interaction $V^{OGE}_{CM}\propto (\vlab_i\cdot\vlab_j)({\vsig}_i\cdot{\vsig}_j)$ which provides mass splittings. The form of $V^{Conf}_{ij}(r_{ij})$ is given by
\begin{equation}\label{vconf}
V^{Conf}_{ij}(r_{ij})=-\frac{3}{16}({\vlab}_i\cdot{\vlab}_j)\cdot b r_{ij},
\end{equation}
where the parameter $b$ denotes the strength of the confinement potential.

There are nine parameters $m_c$, $m_b$, $\alpha_{cc}$, $\alpha_{bb}$, $\alpha_{bc}$, $\sigma_{cc}$, $\sigma_{bb}$, $\sigma_{bc}$, and $b$ to be determined in the calculations. In Refs.~\cite{Deng:2016stx,Li:2019tbn}, the masses of $c\bar{c}$ and $b\bar{c}$ spectrum are calculated by using the three-point difference central method~\cite{cai:2003ktl} from the center ($r=0$) towards outside ($r\to\infty$) point by point. The parameters $m_c$, $\alpha_{cc}$, $\sigma_{cc}$, $b$, $m_b$, $\alpha_{bc}$, $\sigma_{bc}$ have been determined. In this work, we use the same method to determine the parameters $\alpha_{bb}$, $\sigma_{bb}$, by fitting the masses of $b\bar{b}$ spectrum. The parameter set is listed in Table~\ref{parameters}. The corresponding theoretical results for the masses of heavy quarkonia $b\bar{b}$ are shown in Table~\ref{meson mass}.


\subsection{ Matrix elements in color and spin spaces}{\label{M-sfc}}

In order to obtain the mass of a tetraquark state from the nonrelativistic Hamiltonian defined in Eq.~(\ref{Hamiltonian}),
first one needs to calculate the matrix elements of $\langle\vlab_i\cdot\vlab_j\rangle$ and
$\langle\vsig_i\cdot\vsig_j\rangle$ in the color and spin spaces, respectively.

In the color space, one has two kinds of a color-singlet state,
\begin{flalign}
\zeta_1=|6\bar{6}\rangle=|(Q_1Q_2)^6(\bar{Q}_3\bar{Q}_4)^{\bar{6}}\rangle^0,\\
\zeta_2=|\bar{3}3\rangle=|(Q_1Q_2)^{\bar{3}}(\bar{Q}_3\bar{Q}_4)^3\rangle^0.
\end{flalign}
According to the SU(3) Clebsch-Gordan coefficients, one easily obtains the expressions of the color wave functions as follows~\cite{Liu:2004kd,Liu:2016ogz,deSwart:1963pdg,Kaeding:1995vq}:
\begin{equation}\label{colourf66}
\begin{split}
\zeta_1=&\frac{1}{2\sqrt{6}}\bigg[(rb+br)(\bar{b\mathstrut}\bar{r\mathstrut}+\bar{r\mathstrut}\bar{b\mathstrut})+(gr+rg)(\bar{g\mathstrut}\bar{r\mathstrut}+\bar{r\mathstrut}\bar{g\mathstrut})\\
&+(gb+bg)(\bar{b\mathstrut}\bar{g\mathstrut}+\bar{g\mathstrut}\bar{b\mathstrut})\\
&+2(rr)(\bar{r}\bar{r})+2(gg)(\bar{g}\bar{g})+2(bb)(\bar{b}\bar{b})\bigg],
\end{split}
\end{equation}
\begin{equation}\label{colourf33}
\begin{split}
\zeta_2=&\frac{1}{2\sqrt{3}}\bigg[(br-rb)(\bar{b\mathstrut}\bar{r\mathstrut}-\bar{r\mathstrut}\bar{b\mathstrut})-(rg-gr)(\bar{g\mathstrut}\bar{r\mathstrut}-\bar{r\mathstrut}\bar{g\mathstrut})\\
&+(bg-gb)(\bar{b\mathstrut}\bar{g\mathstrut}-\bar{g\mathstrut}\bar{b\mathstrut})\bigg],
\end{split}
\end{equation}
with these color wave functions one can work out the matrix elements $\langle\vlab_i\cdot\vlab_j\rangle$~\cite{Vijande:2009ac},
which have been summarized in Table~\ref{color matrix}.

\begin{table}[htp]
\begin{center}
\caption{\label{color matrix} Color matrix elements.}
\scalebox{1.0}{
\begin{tabular}{cccccccccccc}\hline\hline
&     &$\langle{\vlab_1}\cdot{\vlab_2}\rangle$   &~$\langle{\vlab_3}\cdot{\vlab_4}\rangle$   &$\langle{\vlab_1}\cdot{\vlab_3}\rangle$   &$\langle{\vlab_2}\cdot{\vlab_4}\rangle$   &~$\langle{\vlab_1}\cdot{\vlab_4}\rangle$   &$\langle{\vlab_2}\cdot{\vlab_3}\rangle$\\
\hline
&  $\langle\zeta_1|\hat{O}|\zeta_1\rangle$     &~4/3                                       &~4/3                                       &~$-10/3$
&~$-10/3$                                     &~$-10/3$                                     &~$-10/3$                                   ~~\\
&  $\langle\zeta_2|\hat{O}|\zeta_2\rangle$     &~$-8/3$                                      &~$-8/3$                                      &~$-4/3$
&~$-4/3$                                      &~$-4/3$                                      &~$-4/3$                                    ~~\\
&  $\langle \zeta_1|\hat{O}|\zeta_2\rangle$     &~0                                         &~~0                                        &~$-2\sqrt{2}$
&~$-2\sqrt{2}$                              &~$2\sqrt{2}$                               &~$2\sqrt{2}$                              ~~\\
\hline\hline
\end{tabular}}
\end{center}
\end{table}

In the spin space, one has six spin wave functions,
\begin{flalign}
\chi_0^{00}=|(Q_1Q_2)_0(\bar{Q}_3\bar{Q}_4)_0\rangle_0,\\
\chi_0^{11}=|(Q_1Q_2)_1(\bar{Q}_3\bar{Q}_4)_1\rangle_0,\\
\chi_1^{01}=|(Q_1Q_2)_0(\bar{Q}_3\bar{Q}_4)_1\rangle_1,\\
\chi_1^{10}=|(Q_1Q_2)_1(\bar{Q}_3\bar{Q}_4)_0\rangle_1,\\
\chi_1^{11}=|(Q_1Q_2)_1(\bar{Q}_3\bar{Q}_4)_1\rangle_1,\\
\chi_2^{11}=|(Q_1Q_2)_1(\bar{Q}_3\bar{Q}_4)_1\rangle_2.
\end{flalign}
According to the SU(2) Clebsch-Gordan coefficients, we easily obtain the expressions of the spin wave functions as follows:
\begin{eqnarray}\label{spin000}
\chi_0^{00}&=&\frac{1}{2}(\uparrow\downarrow\uparrow\downarrow-\uparrow\downarrow\downarrow\uparrow-\downarrow\uparrow\uparrow\downarrow+\downarrow\uparrow\downarrow\uparrow),\\
\chi_0^{11}&=&\sqrt{\frac{1}{12}}(2\uparrow\uparrow\downarrow\downarrow-\uparrow\downarrow\uparrow\downarrow
-\uparrow\downarrow\downarrow\uparrow\nonumber \\
&&-\downarrow\uparrow\uparrow\downarrow-
\downarrow\uparrow\downarrow\uparrow+2\downarrow\downarrow\uparrow\uparrow),\\
\chi_1^{01}&=&\sqrt{\frac{1}{2}}(\uparrow\downarrow\uparrow\uparrow-\downarrow\uparrow\uparrow\uparrow),\\
\chi_1^{10}&=&\sqrt{\frac{1}{2}}(\uparrow\uparrow\uparrow\downarrow-\uparrow\uparrow\downarrow\uparrow),\\
\chi_1^{11}&=&\frac{1}{2}(\uparrow\uparrow\uparrow\downarrow+\uparrow\uparrow\downarrow\uparrow
-\uparrow\downarrow\uparrow\uparrow-\downarrow\uparrow\uparrow\uparrow),\\
\chi_2^{11}&=&\uparrow\uparrow\uparrow\uparrow,
\end{eqnarray}
with these spin wave functions one can work out the matrix elements of $\langle\vsig_i\cdot\vsig_j\rangle$~\cite{Vijande:2009ac},
which have been listed in Table~\ref{spin matrix}.

\begin{table*}[htp]
\begin{center}
\caption{\label{spin matrix} Spin matrix elements.}{
\begin{tabular}{cccccccccccc}\hline\hline
&                                              &~~~~~$\langle{\vsig_1}\cdot{\vsig_2}\rangle$~~~~~   &~~~~~$\langle{\vsig_3}\cdot{\vsig_4}\rangle$~~~~~   &~~~~~$\langle{\vsig_1}\cdot{\vsig_3}\rangle$~~~~~   &~~~~~$\langle{\vsig_2}\cdot{\vsig_4}\rangle$~~~~~   &~~~~~$\langle{\vsig_1}\cdot{\vsig_4}\rangle$~~~~~   &~~~~~$\langle{\vsig_2}\cdot{\vsig_3}\rangle$~~~~~   \\
\hline
&  $\langle\chi_0^{00}|\hat{O}|\chi_0^{00}\rangle$          &~-3                                          &~-3                                          &~0
&~0                                           &~0                                           &~0                                                     \\
&  $\langle\chi_0^{11}|\hat{O}|\chi_0^{11}\rangle$          &~1                                           &~1                                           &~-2
&~-2                                          &~-2                                          &~-2                                           \\
&  $\langle\chi_0^{00}|\hat{O}|\chi_0^{11}\rangle$          &~0                                           &~~0                                          &~$-\sqrt{3}$
&~$-\sqrt{3}$                                 &~$\sqrt{3}$                                  &~$\sqrt{3}$                                 \\
&  $\langle\chi_1^{01}|\hat{O}|\chi_1^{01}\rangle$          &~-3                                          &~1                                           &~0
&~0                                           &~0                                           &~0                                            \\
&  $\langle\chi_1^{10}|\hat{O}|\chi_1^{10}\rangle$          &~1                                           &~-3                                          &~0
&~0                                           &~0                                           &~0                                          \\
&  $\langle\chi_1^{11}|\hat{O}|\chi_1^{11}\rangle$          &~1                                           &~~1                                          &~-1
&~-1                                          &~-1                                          &~-1                                          \\
&  $\langle\chi_1^{01}|\hat{O}|\chi_1^{10}\rangle$          &~0                                           &~~0                                          &~1
&~1                                           &~-1                                          &~-1                                          \\
&  $\langle\chi_1^{01}|\hat{O}|\chi_1^{11}\rangle$          &~0                                           &~~0                                          &~$-\sqrt{2}$
&~$\sqrt{2}$                                  &~$-\sqrt{2}$                                 &$\sqrt{2}$                                    \\
&  $\langle\chi_1^{10}|\hat{O}|\chi_1^{11}\rangle$          &~0                                           &~~0                                          &~$\sqrt{2}$
&~$-\sqrt{2}$                                 &~$-\sqrt{2}$                                 &~$\sqrt{2}$                                  \\
&  $\langle\chi_2^{11}|\hat{O}|\chi_2^{11}\rangle$          &~1                                           &~~1                                          &~1
&~1                                           &1                                            &1                                            \\
\hline\hline
\end{tabular}}
\end{center}
\end{table*}

\subsection{ Matrix elements in the coordinate space}{\label{M-sfc}}

The trail wave function of the four-quark states without any spatial excitations in the coordinate space is
expanded by a series of Gaussian functions,
\begin{equation}\label{cf1}
\psi({\mathbf{r_1},\mathbf{r_2},\mathbf{r_3},\mathbf{r_4}})=\prod_{i=1}^4\sum_{\ell=1}^n\mathcal{C}_{i\ell}\left(\frac{1}{\pi b^2_{i\ell}}\right)^{3/4}\exp\left[-\frac{1}{2 b^2_{i\ell}}r^2_i\right],
\end{equation}
where the parameters $b_{i\ell}$ are related to the harmonic oscillator frequencies $\omega_\ell$ with $1/b^2_{i\ell}=m_i\omega_\ell$.
With the ansatz of the mass independent frequency $\omega_\ell$ for a quark of mass $m_i$~\cite{Straub:1988gj}, i.e.,
$1/b^2_{i\ell}=1/b^2_\ell=m_u \omega_\ell$ ($m_u$=313 MeV), the trail wave function of the four quark states can be simplified to be
\begin{eqnarray}\label{gsfa}
\psi({\mathbf{r_1},\mathbf{r_2},\mathbf{r_3},\mathbf{r_4}})&=&\sum^n_{\ell} \mathcal{C}_{\ell}\prod_{i=1}^4\left(\frac{m_i\omega_{\ell}}{\pi}\right)^{3/4}\exp\left[-\frac{m_i\omega_l}{2}r^2_i\right]\nonumber\\
&\equiv&\sum^n_{\ell} \mathcal{C}_{\ell}\prod_{i=1}^4 \phi(\omega_\ell,\mathbf{r}_i),
\end{eqnarray}
which is often adopted in the calculations of the multiquark systems~\cite{Zhang:2007mu,Zhang:2005jz}.

In the coordinate space, we need work out the matrix elements of $\langle1/r_{ij}\rangle$, $\langle e^{-\sigma^2_{ij}r_{ij}^2}\rangle$, and $\langle r_{ij}\rangle$. Combing the trail wave functions defined in Eq.~(\ref{gsfa}), we obtain
\begin{equation}\label{r matrix element}
\left\langle\psi(\omega_\ell,\mathbf r_{i}, \mathbf r_{j})\left|\frac{1}{r_{ij}}\right|\psi(\omega_{\ell'},\mathbf r_{i}, \mathbf r_{j})\right\rangle=2\sqrt{\frac{m_{ij}}{\pi}}\frac{(\omega_\ell\omega_{\ell'})^{3/2}}{(\frac{\omega_\ell+\omega_{\ell'}}{2})^{5/2}},
\end{equation}
\begin{equation}\label{sigp matrix element}
\left\langle\psi(\omega_{\ell},\mathbf r_{i}, \mathbf r_{j})\left|e^{-\sigma^2_{ij}r_{ij}^2}\right|\psi(\omega_{\ell'},\mathbf r_{i}, \mathbf r_{j})\right\rangle
=\left(\frac{m_{ij}(\frac{2\omega_\ell\omega_{\ell'}}{\omega_\ell+\omega_{\ell'}})}{m_{ij}\frac{\omega_\ell+\omega_{\ell'}}{2}+\sigma_{ij}^2}\right)^{\frac{3}{2}},
\end{equation}
\begin{equation}\label{sigp matrix element}
\left\langle\psi(\omega_\ell,\mathbf r_{i}, \mathbf r_{j})\left| r_{ij}\right|\psi(\omega_{\ell'},\mathbf r_{i}, \mathbf r_{j})\right\rangle=2\sqrt{\frac{1}{\pi m_{ij}}}\frac{(\omega_\ell\omega_{\ell'})^{3/2}}{(\frac{\omega_\ell+\omega_{\ell'}}{2})^{7/2}},
\end{equation}
where $\psi(\omega_i,\mathbf r_{i}, \mathbf r_{j})\equiv \phi(\omega_\ell,\mathbf{r}_i)\phi(\omega_\ell,\mathbf{r}_j)$, $m_{ij}=m_im_j/(m_i+m_j)$.

To separate out the center-of-mass kinetic energy $T_G$ and finally work out the kinetic energy matrix element $\langle\sum_{i=1}^4T_i-T_G\rangle$ , we need to redefine the coordinates by the following Jacobi coordinates,
\begin{eqnarray}
\vxi_1&\equiv&\mathbf{r_1}-\mathbf{r_2},\\
\vxi_2&\equiv&\mathbf{r_3}-\mathbf{r_4},\\
\vxi_3&\equiv&\frac{m_1\mathbf{r_1}+m_2\mathbf{r_2}}{m_1+m_2}-\frac{m_3\mathbf{r_3}+m_4\mathbf{r_4}}{m_3+m_4},\\
\vxi_{4}&\equiv&\frac{m_1\mathbf{r_1}+m_2\mathbf{r_2}+m_3\mathbf{r_3}+m_4\mathbf{r_4}}{m_1+m_2+m_3+m_4},
\end{eqnarray}
with these one can rewritten the Eq.~(\ref{gsfa}) as
\begin{equation}\label{gsf Jacobia}
\psi(\vxi_1,\vxi_2,\vxi_3,\vxi_4)=\sum^n_\ell \mathcal{C}_{\ell}\prod_{i=1}^4\left(\frac{\mu_i\omega_\ell}{\pi}\right)^{3/4}\exp\left[-\frac{\mu_i\omega_\ell}{2}\xi^2_i\right],
\end{equation}
where $\mu_1\equiv m_1m_2/(m_1+m_2)$, $\mu_2\equiv m_3m_4/(m_3+m_4)$, $\mu_3\equiv (m_1+m_2)(m_3+m_4)/M$, $\mu_4=M\equiv m_1+m_2+m_3+m_4$.
With the trail wave function defined in Eq.~(\ref{gsf Jacobia}), the kinetic energy matrix element is worked out to be
\begin{equation}\label{T Jacobi}
\left\langle\sum_{i=1}^4T_i-T_G\right\rangle
=\frac{9}{4}\sum_{\ell}^{n}\sum_{\ell'}^{n}\mathcal{C}_{\ell}\mathcal{C}_{\ell'}\frac{(\omega_\ell\omega_{\ell'})^4}{(\frac{\omega_\ell+\omega_{\ell'}}{2})^7}.
\end{equation}

\begin{table*}[htp]
\begin{center}
\caption{\label{cccc} Predicted mass spectra for the $cc\bar{c}\bar{c}$, $bb\bar{b}\bar{b}$ and $bb\bar{c}\bar{c}$ systems.}
\scalebox{1.0}{
\begin{tabular}{cccccccccccc}\hline\hline
~& $J^{P(C)}$  ~& Configuration                                             ~& $\langle H\rangle$ (MeV)
                                                                               ~& Mass (MeV)
                                                                               ~& Eigenvector\\
\hline
~& $0^{++}$  ~& $|\{cc\}^6_0\{\bar{c}\bar{c}\}^{\bar{6}}_0\rangle^0_0$    ~& \multirow{2}{*}{$\begin{pmatrix}6518&-45 \\-45&6487\end{pmatrix}$}
                                                                               ~& \multirow{2}{*}{$\begin{bmatrix}6550 \\6455 \end{bmatrix}$}
                                                                               ~& \multirow{2}{*}{$\begin{bmatrix}(0.81, -0.58)\\(0.58, 0.81)\end{bmatrix}$}\\
~&                ~& $|\{cc\}^{\bar{3}}_1\{\bar{c}\bar{c}\}^3_1\rangle^0_0$\\
\\
~& $1^{+-}$  ~& $|\{cc\}^{\bar{3}}_1\{\bar{c}\bar{c}\}^3_1\rangle^0_1$    ~& {$\begin{pmatrix}6500\end{pmatrix}$}
                                                                               ~&  6500
                                                                               ~&  1\\
\\
~& $2^{++}$  ~& $|\{cc\}^{\bar{3}}_1\{\bar{c}\bar{c}\}^3_1\rangle^0_2$    ~& {$\begin{pmatrix}6524\end{pmatrix}$}
                                                                               ~&  6524
                                                                               ~&  1\\
\hline\hline
~& $0^{++}$  ~& $|\{bb\}^6_0\{\bar{b}\bar{b}\}^{\bar{6}}_0\rangle^0_0$    ~& \multirow{2}{*}{$\begin{pmatrix}19338&-23 \\-23 &19322\end{pmatrix}$}
                                                                               ~& \multirow{2}{*}{$\begin{bmatrix}19355\\19306\end{bmatrix}$}
                                                                               ~& \multirow{2}{*}{$\begin{bmatrix}(0.81, -0.58)\\(0.58, 0.81)\end{bmatrix}$}\\
~&                ~& $|\{bb\}^{\bar{3}}_1\{\bar{b}\bar{b}\}^3_1\rangle^0_0$\\
\\
~& $1^{+-}$  ~& $|\{bb\}^{\bar{3}}_1\{\bar{b}\bar{b}\}^3_1\rangle^0_1$    ~& {$\begin{pmatrix}19329\end{pmatrix}$}
                                                                               ~&  19329
                                                                               ~&  1\\
\\
~& $2^{++}$  ~& $|\{bb\}^{\bar{3}}_1\{\bar{b}\bar{b}\}^3_1\rangle^0_2$    ~& {$\begin{pmatrix}19341\end{pmatrix}$}
                                                                               ~&  19341
                                                                               ~&  1\\
\hline\hline
~& $0^{+}$  ~& $|\{bb\}^6_0\{\bar{c}\bar{c}\}^{\bar{6}}_0\rangle^0_0$       ~& \multirow{2}{*}{$\begin{pmatrix}13032&-23\\-23&12953 \end{pmatrix}$}
                                                                               ~& \multirow{2}{*}{$\begin{bmatrix}13039\\12947\end{bmatrix}$}
                                                                               ~& \multirow{2}{*}{$\begin{bmatrix}(0.97, -0.26)\\(0.26, 0.97)\end{bmatrix}$}\\
~&             ~& $|\{bb\}^{\bar{3}}_1\{\bar{c}\bar{c}\}^3_1\rangle^0_0$\\
\\
~& $1^{+}$  ~& $|\{bb\}^{\bar{3}}_1\{\bar{c}\bar{c}\}^3_1\rangle^0_1$       ~& {$\begin{pmatrix}12960\end{pmatrix}$}
                                                                               ~&  12960
                                                                               ~&  1\\
\\
~& $2^{+}$  ~& $|\{bb\}^{\bar{3}}_1\{\bar{c}\bar{c}\}^3_1\rangle^0_2$       ~& {$\begin{pmatrix}12972\end{pmatrix}$}
                                                                               ~&  12972
                                                                               ~&  1\\
\hline\hline
\end{tabular}}
\end{center}
\end{table*}

\begin{table*}[htp]
\begin{center}
\caption{\label{ccmass} Our predicted masses (MeV) for the $cc\bar{c}\bar{c}$ system compared with others.}
\begin{tabular}{c|ccccccccccccc}\hline\hline
State  & Ours  &Ref.~\cite{Wu:2016vtq}    &Ref.~\cite{Lloyd:2003yc}         &Ref.~\cite{Chen:2016jxd}
               &Ref.~\cite{Ader:1981db}   &Ref.~\cite{Iwasaki:1975pv}        &Ref.~\cite{Karliner:2016zzc}     &Ref.~\cite{Barnea:2006sd}
               &Refs.~\cite{Wang:2017jtz,Wang:2018poa}  &Ref.~\cite{Debastiani:2017msn}   &Ref.~\cite{Berezhnoy:2011xn} & Ref.~\cite{Anwar:2017toa}
\\ \hline
$0^{++}$ &6487 &6797 &6477 & 6460-6470 &6437 &6200 &6192 &6038-6115  &5990   &5969 &5966 & $< 6140$ \\
$0^{++}$ &6518 &7016 &6695 & 6440-6820 &6383 &...  &...  &...        &...    &...  &...  &  \\
$1^{+-}$ &6500 &6899 &6528 & 6370-6510 &6437 &...  &...  &6101-6176  &6050    &6021 &6051 & \\
$2^{++}$ &6524 &6956 &6573 & 6370-6510 &6437 &...  &...  &6172-6216  &6090   &6115 &6223 & \\
\hline\hline
\end{tabular}
\end{center}
\end{table*}

\begin{table*}[htp]
\begin{center}
\caption{\label{bbbb other} Our predicted masses (MeV) for the $bb\bar{b}\bar{b}$ system compared with others. }
\begin{tabular}{c|cccccccccccccc}\hline\hline
State  & Ours  &Ref.~\cite{Wu:2016vtq} &Ref.~\cite{Wang:2017jtz,Wang:2018poa} &Ref.~\cite{Karliner:2016zzc}
&Ref.~\cite{Berezhnoy:2011xn} &Ref.~\cite{Anwar:2017toa} &Ref.~\cite{Bai:2016int}         &Ref.~\cite{Chen:2016jxd}
                       &Ref.~\cite{Hughes:2017xie}  &Ref.~\cite{Anwar:2017toa}
\\ \hline

$0^{++}$ &19322 &20155 &18840   &18826 &18754  &18720 &18690 & 18460-18490   &18798      &$< 18890$     \\

$0^{++}$ &19338 &20275 &...     &...   &...    &...   &...   & 18450-19640   &...        &              \\

$1^{+-}$ &19329 &20212 &18840   &...   &18808  &...   &...   & 18320-18540   &...        &              \\

$2^{++}$ &19341 &20243 &18850   &...   &18916  &...   &...   & 18320-18530   &...        &              \\

\hline\hline
\end{tabular}
\end{center}
\end{table*}

\section{results and discussions }\label{Numerical results and discussions}

In this work, we adopt the variation principle to solve the Schr\"{o}dinger equation.
Following the method used in Ref.~\cite{Hiyama:2003cu}, the oscillator length $b_\ell$ are set to be
\begin{equation}\label{geometric progression}
b_\ell=b_1a^{\ell-1}\ \ \ (\ell=1,...,n),
\end{equation}
where $n$ is the number of Gaussian functions, and $a$ is the ratio coefficient. There are three parameters $\{b_1,b_n,n\}$ to be determined through variation method. It is found that when we take $b_1=0.1$ fm, $b_n=4$ fm, $n=15$, we will obtain stable solutions for the four-quark systems.

When all the matrix elements have been worked out, we can solve the generalized matrix eigenvalue problem,
\begin{equation}\label{eigenvalue problem}
\sum_{\ell=1}^{n}\sum_{\ell'=1}^{n}(H_{\ell\ell'}-E_\ell N_{\ell\ell'})\mathcal{C}_{\ell'}^\ell=0,
\end{equation}
where
\begin{equation}\label{Hllp}
H_{\ell\ell'}=\left\langle\psi(\omega_\ell)\varphi \zeta \chi \Big|H\Big|\psi(\omega_{\ell'})\varphi\zeta \chi\right\rangle,
\end{equation}
\begin{equation}\label{Nllp}
N_{\ell\ell'}=\left\langle\psi(\omega_\ell)\varphi \zeta \chi \Big|\psi(\omega_{\ell'})\varphi\zeta \chi\right\rangle,
\end{equation}
with $\psi(\omega_\ell)=\prod_{i=1}^4\left(\frac{\mu_i\omega_\ell}{\pi}\right)^{3/4}\exp\left[-\frac{\mu_i\omega_\ell}{2}\xi^2_i\right]$.
$\varphi$, $\zeta$, and $\chi$ stand for the flavor, color, and spin wave functions, respectively. The physical state corresponds to the solution with a minimum energy $E_{m}$. By solving this generalized matrix eigenvalue problem, the mass of the tetraquark configuration and its spacial wave function can be determined.

\subsection{The $cc\bar{c}\bar{c}$ and $bb\bar{b}\bar{b}$ systems}

The predicted mass spectrum for the $cc\bar{c}\bar{c}$ system has been given in Table~\ref{cccc} and also shown in Fig.~\ref{spectrum} (a).
From Table~\ref{cccc}, it is found that in the two $I^G(J^{PC})=0^+(0^{++})$ states there is a sizable configuration mixing between $|\{cc\}^6_0\{\bar{c}\bar{c}\}^{\bar{6}}_0\rangle^0_0$ and $|\{cc\}^{\bar{3}}_1\{\bar{c}\bar{c}\}^3_1\rangle^0_0$.
The $J^{PC}=0^{++}$ state with a higher mass, 6550 MeV, is dominant by the $|\{cc\}^6_0\{\bar{c}\bar{c}\}^{\bar{6}}_0\rangle^0_0$ configuration, while another $J^{PC}=0^{++}$ state with a lower mass, 6455 MeV, is dominant by the $|\{cc\}^{\bar{3}}_1\{\bar{c}\bar{c}\}^3_1\rangle^0_0$ configuration. The mass splitting between these two $J^{PC}=0^{++}$ states is about $95$ MeV. The other two states $J^{PC}=1^{+-}$  and $J^{PC}=2^{++}$ are also located in a similar mass region, i.e. $\sim 6.5$ GeV, and the mass splitting between them is about 20 MeV. As shown in Fig.~\ref{spectrum} (a), the two $J^{PC}=0^{++}$ states are above the thresholds of the charmonium pairs for about $260\sim 580$ MeV. It suggests that the $J^{PC}=0^{++}$ states are unstable, and they can easily decay into the $\eta_c\eta_c$ and $J/\psi J/\psi$ final states through quark rearrangements. The $J^{PC}=1^{+-}$ state lies about 420 MeV above the mass threshold of $\eta_c J/\psi$, while $J^{PC}=2^{++}$ is about 330 MeV above the mass threshold of $J/\psi J/\psi$, they can also easily decay into $\eta_c J/\psi$ and $J/\psi J/\psi$; respectively, through the quark rearrangements.

As a comparison, our predicted masses and some other typical results from other works are collected in Table~\ref{ccmass}.
It shows that our predicted masses for the $cc\bar{c}\bar{c}$ system are roughly compatible with the nonrelativistic quark model predictions of Refs.~\cite{Lloyd:2003yc,Ader:1981db}, where both confining and Coulomb potentials are considered. It is also interesting to find that similar results are given by the QCD sum rules~\cite{Chen:2016jxd}. In contrast, the masses predicted by us are much larger than those predicted in Refs.~\cite{Berezhnoy:2011xn,Debastiani:2017msn,Wang:2017jtz,Barnea:2006sd,Karliner:2016zzc,Anwar:2017toa}.
These methods which obtained small masses have some common features: either no confining potentials were explicitly included ~\cite{Berezhnoy:2011xn,Barnea:2006sd,Karliner:2016zzc,Anwar:2017toa} or a diquark picture was adopted in the calculations~\cite{Wang:2017jtz,Debastiani:2017msn}. Recently, Wu {\it et al.} also obtained a large mass $\sim6.8-7.0$ GeV for the $cc\bar{c}\bar{c}$ system with the heavier constituent $c$-quark mass $1.72$ GeV adopted~\cite{Wu:2016vtq}.

We further analyze the contributions from each part of the Hamiltonian for the $cc\bar{c}\bar{c}$ system. The results are listed in Table~\ref{cccc xx}. It shows that the averaged kinetic energy $\langle T\rangle$, the confining potential $\langle V^{Conf}\rangle$, and the Coulomb potential $\langle V^{OGE}_{coul}\rangle$ have the same order of magnitude. In particular, the contributions from the confining potential are sizeable and apparently cannot be neglected. Note that the confining potential contributes a positive energy to the system. Thus, neglecting this contribution will lead to much lower masses for the all-heavy system. In Refs.~\cite{Berezhnoy:2011xn,Barnea:2006sd,Karliner:2016zzc,Anwar:2017toa}, the confining potential was explicitly neglected. Although part of the confining potential effects can be taken into account by the effective constituent quark masses in the ground states, our calculation shows that the impact from the inclusion of the confining potential seems not to be on the constituent quark masses in the heavy quark sector, but rather on the relative strengths of the averaged matrix elements among the terms of the nonrelativistic Hamiltonian.

In order to examine the role played by the confining potential in the spectrum of heavy quark system, we compare the contributions from the OGE and confining potential for the $\eta_c$ meson, i.e, $\langle V^{OGE}_{coul}\rangle\simeq -637$ MeV and $\langle V^{Conf}\rangle\simeq 233$ MeV,  which are consistent with our previous study in Ref.~\cite{Deng:2016stx}.
The ratio between the confining potential $\langle V^{Conf}\rangle$ and color Coulomb potential $\langle V^{OGE}_{coul}\rangle$ can reach up to
\begin{equation}\label{ccca}
\left|\frac{\langle V^{conf}\rangle}{\langle V^{OGE}_{coul}\rangle}\right|\simeq 36\%.
\end{equation}
This explicit result suggests that the neglect of confining potential cannot be justified for the $c\bar{c}$ system.

As a general conclusion, we find that the confining potential has significant
contributions to the masses of the $cc\bar{c}\bar{c}$ system, and are the same order of magnitude as the color Coulomb potential. This will enhance the masses of the $cc\bar{c}\bar{c}$ system and does not support the existence of a bound tetraquark of $cc\bar{c}\bar{c}$ with narrow widths.

\begin{table}[htp]
\begin{center}
\caption{\label{cccc xx} The contributions from each part of the Hamiltonian of the $cc\bar{c}\bar{c}$ and $bb\bar{b}\bar{b}$ systems in units of MeV.}
\scalebox{1.0}{
\begin{tabular}{cccccccccccc}\hline\hline
~& $J^{PC}$  ~& Configuration  ~& $M$          ~        ~& $\langle T\rangle$        ~& $\langle V^{Conf}\rangle$    ~& $\langle V^{OGE}_{coul}\rangle$    ~& $\langle V^{OGE}_{CM}\rangle$\\
\hline
~& $0^{++}$  ~& $|\{cc\}^6_0\{\bar{c}\bar{c}\}^{\bar{6}}_0\rangle^0_0$
                                    ~& 6518          ~& 715      ~& 664         ~& $-811$              ~& 18 \\

~&                ~& $|\{cc\}^{\bar{3}}_1\{\bar{c}\bar{c}\}^3_1\rangle^0_0$
                                    ~& 6487           ~& 756     ~& 646        ~& $-834 $             ~& $-13$\\
\\
~& $1^{+-}$  ~& $|\{cc\}^{\bar{3}}_1\{\bar{c}\bar{c}\}^3_1\rangle^0_1$
                                    ~& 6500            ~& 739      ~& 653         ~& $-825$              ~& 0\\
\\
~& $2^{++}$  ~& $|\{cc\}^{\bar{3}}_1\{\bar{c}\bar{c}\}^3_1\rangle^0_2$
                                    ~& 6524       ~& 708      ~& 667         ~& $-806$             ~& 23\\
\hline
~& $0^{++}$  ~& $|\{bb\}^6_0\{\bar{b}\bar{b}\}^{\bar{6}}_0\rangle^0_0$
                                    ~& 19338           ~& 768      ~& 356        ~& $-1203$             ~& 9 \\

~&                ~& $|\{bb\}^{\bar{3}}_1\{\bar{b}\bar{b}\}^3_1\rangle^0_0$
                                    ~& 19322           ~& 796      ~& 350        ~& $-1225$             ~& $-6$\\
\\
~& $1^{+-}$  ~& $|\{bb\}^{\bar{3}}_1\{\bar{b}\bar{b}\}^3_1\rangle^0_1$
                                    ~& 19329           ~& 785      ~& 353        ~& $-1216$            ~& 0\\
\\
~& $2^{++}$  ~& $|\{bb\}^{\bar{3}}_1\{\bar{b}\bar{b}\}^3_1\rangle^0_2$
                                    ~& 19341          ~& 763     ~& 357        ~& $-1199$              ~& 12\\  \hline\hline
\end{tabular}}
\end{center}
\end{table}


The predicted mass spectrum for the $bb\bar{b}\bar{b}$ system is very similar to that for the $cc\bar{c}\bar{c}$ one. The results are given in Table~\ref{cccc} and shown in Fig.~\ref{spectrum} (b). The configuration mixing effects between $|\{bb\}^6_0\{\bar{b}\bar{b}\}^{\bar{6}}_0\rangle^0_0$ and $|\{bb\}^{\bar{3}}_1\{\bar{b}\bar{b}\}^3_1\rangle^0_0$ should be obvious in the two $J^{PC}=0^{++}$ states, the higher mass state with mass 19355 MeV is dominant by the $|\{bb\}^6_0\{\bar{b}\bar{b}\}^{\bar{6}}_0\rangle^0_0$ configuration, and the lower one of 19306 MeV is dominant by the $|\{bb\}^{\bar{3}}_1\{\bar{b}\bar{b}\}^3_1\rangle^0_0$ configuration. Due to the heavier mass of the $b$ quark, relatively smaller mass splittings among these states are found. The pattern is also similar to that of the  $cc\bar{c}\bar{c}$ system. Note that the predicted masses are above the thresholds of the bottomonium pairs for about $380\sim 560$ MeV. It suggests that bound states of the $bb\bar{b}\bar{b}$ system with narrow widths are not favored.

In Table~\ref{bbbb other} we compare our results with other model calculations. It shows that our predicted masses are higher than most of the other predictions which are either calculated without including the confining potential explicitly ~\cite{Berezhnoy:2011xn,Barnea:2006sd,Karliner:2016zzc,Anwar:2017toa}, or based on the diquark picture~\cite{Wang:2017jtz}. Similarly, based on the diquark picture, the lightest mass of $bb\bar{b}\bar{b}$ is estimated at $18.8$ GeV by Ref.~\cite{Esposito:2018cwh}. In these calculations the tetraquark states of $J^{PC}=0^{++}$, $1^{+-}$, or $2^{++}$ are either below or slightly above the thresholds of $\eta_b\eta_b$, $\eta_b \Upsilon(1S)$ or $\Upsilon(1S) \Upsilon(1S)$, respectively. Thus, they can become stable with narrow decay widths.  In contrast, our calculations with the inclusion of the confining potential result in higher masses for the $bb\bar{b}\bar{b}$ system and do not favor the existence of such narrow tetraquark states. We note that a rather large mass $\sim20.2$ GeV for the $bb\bar{b}\bar{b}$ system is estimated by Ref.~\cite{Wu:2016vtq}, where a heavier constituent $b$-quark mass $5.05$ GeV is adopted.

In Table~\ref{cccc xx}, the contributions from each part of the Hamiltonian for the $bb\bar{b}\bar{b}$ system are listed. It shows that the kinetic energy $\langle T \rangle\simeq 800$ MeV, the confining potential $\langle V^{Conf}_{ij}(r_{ij})\rangle \simeq 400$ MeV, and the coulomb potential $\langle V^{OGE}_{coul}\rangle\simeq -1200$ MeV, have the same order of magnitude. As shown in Fig.~\ref{spectrum} (b), the mass splittings among these $J^{PC}=0^{++}$, $1^{+-}$, and $2^{++}$ states follow a similar pattern as in the $cc\bar{c}\bar{c}$ system. Also similar to that for the $cc\bar{c}\bar{c}$ system, the neglect of the confining potential will lead to much lower masses for the $bb\bar{b}\bar{b}$ system, and this may explain the low masses obtained in Refs.~\cite{Berezhnoy:2011xn,Barnea:2006sd,Karliner:2016zzc,Anwar:2017toa}.
Although it is often argued that the confining potential contributions are perturbative for the bottomonium system, explicit calculations seem not to support this phenomenon.
In Ref.~\cite{Deng:2016stx}, we have studied the $b\bar{b}$ spectrum and find that $\langle V^{Conf}(r)\rangle\simeq 122$ MeV and $\langle V^{OGE}_{coul}\rangle\simeq -970$ MeV for the $\eta_b$ meson. The ratio between $\langle V^{Conf}\rangle$ and $\langle V^{OGE}_{coul}\rangle$ can reach up to
\begin{equation}\label{bbba}
\left|\frac{\langle V^{conf}\rangle}{\langle V^{OGE}_{coul}\rangle}\right|\simeq 13\%.
\end{equation}
For the four heavy quark system of $bb\bar{b}\bar{b}$, the increase of the displacements between the two quarks (antiquarks) or quark-antiquark will experience larger confining forces. Thus, the confining potential contributions cannot be neglected in the calculations. As a consequence, our study does not support the existence of the tetraquark $bb\bar{b}\bar{b}$ bound states with narrow widths.

Finally, it should be mentioned that for a simplicity, in our calculation, the variational wave functions of the coordinate space are only adopted an $s$-wave form. Thus, the color wave functions for the $J^{PC}=1^{+-}$ and $2^{++}$ states is color $\bar{3}3$.
However, the $\bar{3}3$ color wave functions for the $J^{PC}=1^{+-}$ and $2^{++}$ states might slightly mix
with the color $6\bar{6}$ when one considers the orbital excitations in the coordinate space~\cite{Richard:2017vry,Vijande:2009kj,Richard:2018yrm}. With a color mixing effect, the mass of the $J^{PC}=1^{+-}$ and $2^{++}$ states might become slightly lower~\cite{Richard:2017vry,Vijande:2009kj,Richard:2018yrm}, which does not affect our conclusions.

\subsection{The $bb\bar{c}\bar{c}$ system}

The $bb\bar{c}\bar{c}$ system is similar to the $cc\bar{c}\bar{c}$ and $bb\bar{b}\bar{b}$ ones except that it does not have determined $C$ parity, and there is no contributions from the annihilation potential.
The predicted mass spectrum for the $bb\bar{c}\bar{c}$ system is also listed in Table~\ref{cccc} and shown in Fig.~\ref{spectrum} (c).
From Table~\ref{cccc}, a small configuration mixing between $|\{bb\}^6_0\{\bar{c}\bar{c}\}^{\bar{6}}_0\rangle^0_0$ and $|\{bb\}^{\bar{3}}_1\{\bar{c}\bar{c}\}^3_1\rangle^0_0$ can be identified. The higher mass state (13039 MeV) of $J^{P}=0^{+}$ state is dominant by the $|\{cc\}^6_0\{\bar{c}\bar{c}\}^{\bar{6}}_0\rangle^0_0$ configuration, while the lower mass one (12947 MeV) of the same quantum numbers is dominant by the $|\{cc\}^{\bar{3}}_1\{\bar{c}\bar{c}\}^3_1\rangle^0_0$ configuration. The mass splitting between these two $J^{P}=0^{+}$ states is about $92$ MeV. The other two states with $J^{P}=1^{+}$  and $2^{+}$ have a small mass splitting of about 10 MeV and are located around $12.96$ GeV.
The masses predicted by us are about 600 MeV systematically smaller than those predicted in the recent work~\cite{Wu:2016vtq}, where relatively large constituent quark masses for the $b$ quark $5.05$ GeV and $c$ quark $1.72$ GeV are adopted.

As shown in Fig.~\ref{spectrum} (c), all these states are above their lowest open flavor decay channels for about 300 MeV. Therefore, they can decay into the $B_c B_c$, $B_c^* B_c^*$, or $B_c B_c^*$ final states via the quark rearrangement quite easily.

\subsection{The $bc\bar{c}\bar{c}$ and $bc\bar{b}\bar{b}$ systems}

The states of both $bc\bar{c}\bar{c}$ and $bc\bar{b}\bar{b}$ systems do not have determined $C$ parity and they share some common features in terms of heavy quark symmetry. The predicted mass spectra for these two configurations are listed in Table~\ref{bccc} and shown in Fig.~\ref{spectrum} (d) and Fig.~\ref{spectrum} (e), respectively. It shows that both $bc\bar{c}\bar{c}$ and $bc\bar{b}\bar{b}$ systems have sizeable configuration mixings between the color $6\otimes\bar{6}$ and $3\otimes\bar{3}$ configurations. For the $bc\bar{c}\bar{c}$ system the mixing occurs between the $|\{bc\}^6_0\{\bar{c}\bar{c}\}^{\bar{6}}_0\rangle^0_0$ and $|\{bc\}^{\bar{3}}_1\{\bar{c}\bar{c}\}^3_1\rangle^0_0$ configurations. It shows that the higher and lower mass states of $J^{P}=0^{+}$ are dominated by the $|\{bc\}^6_0\{\bar{c}\bar{c}\}^{\bar{6}}_0\rangle^0_0$ and $|\{bc\}^{\bar{3}}_1\{\bar{c}\bar{c}\}^3_1\rangle^0_0$ configuration, respectively.

The configuration mixing effects among these three $J^{P}=1^{+}$ states are also sizeable which are shown in Table~\ref{bccc}. The typical mass splitting is about 20 MeV and the predicted masses are about 500 MeV systematically smaller than those predicted in the recent work~\cite{Wu:2016vtq}. Again, we note that rather large constituent quark masses for the $c$ and $b$ quarks are adopted in Ref.~\cite{Wu:2016vtq}.

As a consequence of the high masses predicted by our model, namely, the states of $bc\bar{c}\bar{c}$ system are about $290-350$ MeV above the mass threshold of $B_c^* J/\psi$, we find that these states can easily decay into the $B_c \eta_c$,  $B_c J/\psi$ or $B_c^* J/\psi$ final states via the quark rearrangements. Thus, we do not expect narrow states of $bc\bar{c}\bar{c}$ to be observed in experiment.
\begin{table*}[htp]
\begin{center}
\caption{\label{bccc} Predicted mass spectra for the $bc\bar{c}\bar{c}$ and $bc\bar{b}\bar{b}$ systems. }
\scalebox{1.0}{
\begin{tabular}{cccccccccccc}\hline\hline
~& $J^{P}$  ~& Configuration                                                ~& $\langle H\rangle$ (MeV)
                                                                               ~& Mass (MeV)
                                                                               ~& Eigenvector\\
\hline
~& $0^{+}$  ~& $|(bc)^6_0\{\bar{c}\bar{c}\}^{\bar{6}}_0\rangle^0_0$         ~& \multirow{2}{*}{$\begin{pmatrix}9763 &-34\\-34&9740\end{pmatrix}$}
                                                                               ~& \multirow{2}{*}{$\begin{bmatrix}9787 \\9715\end{bmatrix}$}
                                                                               ~& \multirow{2}{*}{$\begin{bmatrix}(0.81 , -0.58 )\\(0.58 , 0.81 )\end{bmatrix}$}\\
~&             ~& $|(bc)^{\bar{3}}_1\{\bar{c}\bar{c}\}^3_1\rangle^0_0$\\
\\
~& $1^{+}$  ~& $|(bc)^6_1\{\bar{c}\bar{c}\}^{\bar{6}}_0\rangle^0_1$         ~& \multirow{3}{*}{$\begin{pmatrix}9757&-9 &20 \\-9&9749 &4\\20 &4 &9746 \end{pmatrix}$}
                                                                               ~& \multirow{3}{*}{$\begin{bmatrix}9773 \\9752 \\9727\end{bmatrix}$}
                                                                               ~& \multirow{3}{*}{$\begin{bmatrix}(0.80,-0.21,0.56)\\(-0.07,0.90,0.43)\\(-0.59,-0.38,0.71)\end{bmatrix}$}\\
~&             ~& $|(bc)^{\bar{3}}_1\{\bar{c}\bar{c}\}^3_1\rangle^0_1$\\
~&             ~& $|(bc)^{\bar{3}}_0\{\bar{c}\bar{c}\}^3_1\rangle^0_1$\\
\\
~& $2^{+}$  ~& $|(bc)^{\bar{3}}_1\{\bar{c}\bar{c}\}^3_1\rangle^0_2$         ~& {$\begin{pmatrix}9768 \end{pmatrix}$}
                                                                               ~&  9768
                                                                               ~&  1\\
\hline\hline
~& $0^{+}$  ~& $|(bc)^6_0\{\bar{b}\bar{b}\}^{\bar{6}}_0\rangle^0_0$         ~& \multirow{2}{*}{$\begin{pmatrix}16173 &-23\\-23&16158\end{pmatrix}$}
                                                                               ~& \multirow{2}{*}{$\begin{bmatrix}16190 \\16141 \end{bmatrix}$}
                                                                               ~& \multirow{2}{*}{$\begin{bmatrix}(0.81 , -0.58 )\\(0.58 , 0.81 )\end{bmatrix}$}\\
~&             ~& $|(bc)^{\bar{3}}_1\{\bar{b}\bar{b}\}^3_1\rangle^0_0$\\
\\
~& $1^{+}$  ~& $|(bc)^6_1\{\bar{b}\bar{b}\}^{\bar{6}}_0\rangle^0_1$         ~& \multirow{3}{*}{$\begin{pmatrix}16167&0.85&13.40\\0.85&16164&-0.40\\13.40&-0.40&16157\end{pmatrix}$}
                                                                               ~& \multirow{3}{*}{$\begin{bmatrix}16176 \\16164\\16148\end{bmatrix}$}
                                                                               ~& \multirow{3}{*}{$\begin{bmatrix}(0.82, 0.04,0.57)\\(0, 1,-0.06)\\(-0.57, 0.05,0.82)\end{bmatrix}$}\\
~&             ~& $|(bc)^{\bar{3}}_1\{\bar{b}\bar{b}\}^3_1\rangle^0_1$\\
~&             ~& $|(bc)^{\bar{3}}_0\{\bar{b}\bar{b}\}^3_1\rangle^0_1$\\
\\
~& $2^{+}$  ~& $|(bc)^{\bar{3}}_1\{\bar{b}\bar{b}\}^3_1\rangle^0_2$         ~& {$\begin{pmatrix}16176\end{pmatrix}$}
                                                                               ~&  16176
                                                                               ~&  1\\
\hline\hline
\end{tabular}}
\end{center}
\end{table*}


For the $bc\bar{b}\bar{b}$ system its main properties is very similar to that of the $bc\bar{c}\bar{c}$ system as shown in Table~\ref{bccc} and Fig.~\ref{spectrum} (e). Instead of repeating the features seen in the $bc\bar{c}\bar{c}$ system, we only note the main features arising from the heavy constituent quark masses. Namely, the mass splittings among the multiplets with the same quantum numbers are expected to be smaller than that for the $bc\bar{c}\bar{c}$ system. For instance, the mass splitting among the $J^{P}=1^{+}$ states is about 10 MeV.

As shown in Fig.~\ref{spectrum} (e), our results show that the states of the $bc\bar{b}\bar{b}$ system are about $350-390$ MeV above the mass threshold
of $B_c^* \Upsilon$. Thus, these states with different quantum numbers can also easily decay into the $B_c \eta_b$,  $B_c \Upsilon$ or $B_c^* \Upsilon$ final states via the quark rearrangement. Narrow states made of the $bc\bar{b}\bar{b}$ are not favored in our model.

\subsection{The $bc\bar{b}\bar{c}$ system}

\begin{figure*}[htbp]
\begin{center}
\centering  \epsfxsize=18cm \epsfbox{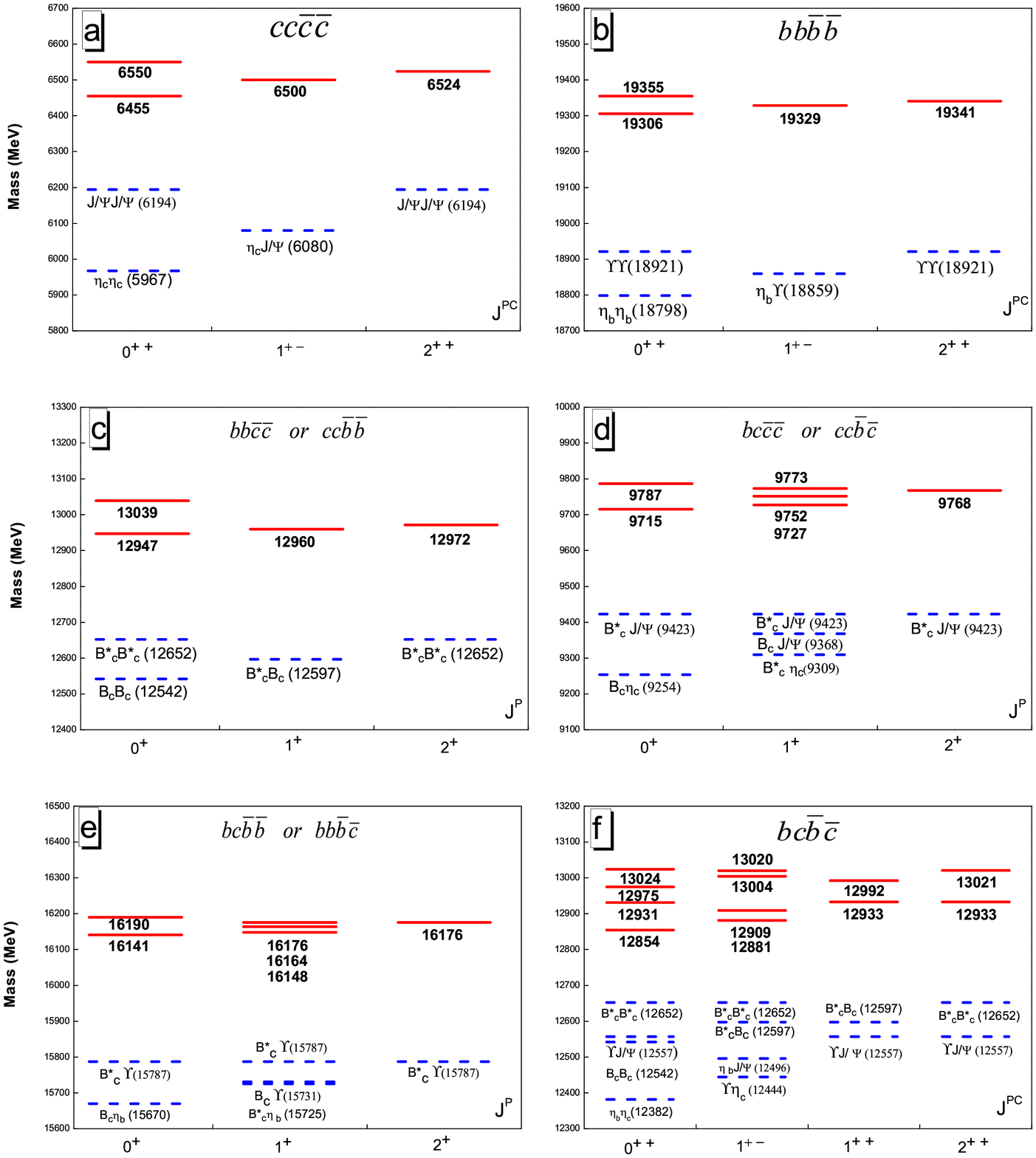}
\vspace{-3cm}\caption{Mass spectra of all-heavy tetraquarks (solid lines) and their possible main decay channels (dashed lines).
The unit of mass is MeV.} \label{spectrum}
\end{center}
\end{figure*}

The $bc\bar{b}\bar{c}$ system has no constraints from the Pauli principle, and there are 12 different configurations allowed by this system, namely,
four $J^{PC}=0^{++}$ states, four $J^{PC}=1^{+-}$ states, two $J^{PC}=1^{++}$ states, and two $J^{PC}=2^{++}$ states. The predicted mass spectrum is listed in Table~\ref{bcbc} and shown in Fig.~\ref{spectrum} (f).

A main feature of the $bc\bar{b}\bar{c}$ system is that the configuration mixing appears to play an important role. For example, the highest mass $J^{PC}=0^{++}$ state is a mixed state containing comparable components from three configurations $|(bc)^6_0(\bar{b}\bar{c})^{\bar{6}}_0\rangle^0_0$, $|(bc)^{\bar{3}}_1(\bar{b}\bar{c})^3_1\rangle^0_0$ and $|(bc)^{\bar{3}}_0(\bar{b}\bar{c})^3_0\rangle^0_0$. As a consequence, the predicted masses for these tetraquark states are in the range of $12939\pm 85$ MeV. We note that our predicted masses are about 600 MeV systematically smaller than those predicted in Ref.~\cite{Wu:2016vtq} and about $350-600$ MeV systematically larger than those predicted with diquark picture in Ref.~\cite{Berezhnoy:2011xn}. Also, these states of $bc\bar{b}\bar{c}$  are about $200-300$ MeV above the mass threshold
of $B_c^* B_c^*$. It suggests that these tetraquark states may easily decay into the $B_c B_c^*$,  $B_c^* B_c^*$, $\eta_b J/\psi$, $\eta_b \eta_c$, $\Upsilon \eta_c$, or $\Upsilon J/\psi$ channels via quark rearrangements. Thus, they are expected to be broad in width.

\begin{table*}[htp]
\begin{center}
\caption{\label{bcbc} Predicted mass spectra for the $bc\bar{b}\bar{c}$ system.}
\scalebox{1.0}{
\begin{tabular}{cccccccccccc}\hline\hline
~& $J^{PC}$  ~& Configuration                                             ~& $\langle H\rangle$ (MeV)
                                                                               ~& Mass (MeV)
                                                                               ~& Eigenvector\\
\hline
~& $0^{++}$  ~& $|(bc)^6_1(\bar{b}\bar{c})^{\bar{6}}_1\rangle^0_0$        ~&\multirow{4}{*}{$\begin{pmatrix}12901&-6&-51&-31\\-6&12956&-29&-44\\
                                                                                                                  -51& -29&12968&-3\\-31&-44&-3&12958\end{pmatrix}$}
                                                                               ~& \multirow{4}{*}{$\begin{bmatrix}12854\\12931\\12975\\13024\end{bmatrix}$}
                                                                               ~& \multirow{4}{*}{$\begin{bmatrix}(0.75, 0.33, 0.43, 0.38)\\(-0.49, 0.68, -0.12, 0.53)\\
                                                                                                                  (-0.27, 0.37, 0.69, -0.57)\\(-0.34, -0.54, 0.58,0.51)
                                                                                                                   \end{bmatrix}$}\\
~&                ~& $|(bc)^6_0(\bar{b}\bar{c})^{\bar{6}}_0\rangle^0_0$\\
~&                ~& $|(bc)^{\bar{3}}_1(\bar{b}\bar{c})^3_1\rangle^0_0$\\
~&                ~& $|(bc)^{\bar{3}}_0(\bar{b}\bar{c})^3_0\rangle^0_0$\\
\\
~& $1^{+-}$  ~& $|(bc)^6_1(\bar{b}\bar{c})^{\bar{6}}_1\rangle^0_1$        ~&\multirow{4}{*}{$\begin{pmatrix}12923&-9&-48&-7\\-9&12946&-7&-61\\
                                                                                                                  -48& -7&12976&-3\\-7&-61&-3&12970\end{pmatrix}$}
                                                                               ~& \multirow{4}{*}{$\begin{bmatrix}12881\\12909\\13004\\13020\end{bmatrix}$}
                                                                               ~& \multirow{4}{*}{$\begin{bmatrix}(0.62, 0.54, 0.37, 0.44)\\(0.61, -0.55, 0.35, -0.46)\\
                                                                                                                  (-0.50, 0.06, 0.86, -0.09)\\(-0.05, -0.63, 0.10,0.77)
                                                                                                                   \end{bmatrix}$}\\
~&                ~& $\frac{1}{\sqrt{2}}|(bc)^6_1(\bar{b}\bar{c})^{\bar{6}}_0\rangle^0_1-|(bc)^6_0(\bar{b}\bar{c})^{\bar{6}}_1\rangle^0_1$\\
~&                ~& $|(bc)^{\bar{3}}_1(\bar{b}\bar{c})^3_1\rangle^0_1$\\
~&                ~& $\frac{1}{\sqrt{2}}|(bc)^{\bar{3}}_1(\bar{b}\bar{c})^3_0\rangle^0_1-|(bc)^{\bar{3}}_0(\bar{b}\bar{c})^3_1\rangle^0_1$\\
\\
~& $1^{++}$  ~& $\frac{1}{\sqrt{2}}|(bc)^6_1(\bar{b}\bar{c})^{\bar{6}}_0\rangle^0_1+|(bc)^6_0(\bar{b}\bar{c})^{\bar{6}}_1\rangle^0_1$
                                                                               ~& \multirow{2}{*}{$\begin{pmatrix}12953&-28\\-28&12973\end{pmatrix}$}
                                                                               ~& \multirow{2}{*}{$\begin{bmatrix}12933\\12992\end{bmatrix}$}
                                                                               ~& \multirow{2}{*}{$\begin{bmatrix}(0.82, 0.58)\\(-0.58, 0.82)\end{bmatrix}$}\\
~&                ~& $\frac{1}{\sqrt{2}}|(bc)^{\bar{3}}_1(\bar{b}\bar{c})^3_0\rangle^0_1+|(bc)^{\bar{3}}_0(\bar{b}\bar{c})^3_1\rangle^0_1$\\
\\
~& $2^{++}$  ~& $|(bc)^6_1(\bar{b}\bar{c})^{\bar{6}}_1\rangle^0_2$        ~& \multirow{2}{*}{$\begin{pmatrix}12962&-41\\-41&12992\end{pmatrix}$}
                                                                               ~& \multirow{2}{*}{$\begin{bmatrix}12933\\13021\end{bmatrix}$}
                                                                               ~& \multirow{2}{*}{$\begin{bmatrix}(0.82, 0.58)\\(-0.58, 0.82)\end{bmatrix}$}\\
~&                ~& $|(bc)^{\bar{3}}_1(\bar{b}\bar{c})^3_1\rangle^0_2$\\
\hline\hline
\end{tabular}}
\end{center}
\end{table*}

\section{Summary}\label{Summary}

In this work, we study the mass spectra of the all-heavy $cc\bar{c}\bar{c}$, $bb\bar{b}\bar{b}$, $bb\bar{c}\bar{c}/cc\bar{b}\bar{b}$, $bc\bar{c}\bar{c}/cc\bar{b}\bar{c}$, $bc\bar{b}\bar{b}/bb\bar{b}\bar{c}$, and $bc\bar{b}\bar{c}$ systems in the potential quark model with the linear confining potential, Coulomb potential, and spin-spin interactions included. We find that the linear confining potential contributes large positive energies to the eigenvalues of the ground states of these tetraquark systems. This is different from some existing calculations in the literature in which the neglect of the confining potential contributions leads to relatively low masses for the all-heavy systems and some of those can be lower than the two-body decay thresholds. In our case, all these states are found to have masses above the corresponding two meson decay thresholds via the quark rearrangement. This implies that narrow all-heavy tetraquark states may not exist in reality. Nevertheless, our explicit calculations suggest that the confining potential still plays an important role in the heavy flavor multiquark system, and it is crucial to include it in dynamical calculations in order to gain a better understanding of the multiquark dynamics. More experimental information from the Belle-II and LHCb analyses would be able to clarify these issues in the near future.

\section*{Acknowledgement}

This work is supported by the National Natural Science Foundation of China (Grants No. 11775078, No. U1832173, No.
11705056, No. 11425525 No. 11521505). Q.Z. is also supported in part, by the DFG and NSFC funds to the Sino-German CRC 110 ``Symmetries and the Emergence of Structure in QCD'' (NSFC Grant No. 11261130311), National Key Basic Research Program of China under Contract No. 2015CB856700.

\bibliographystyle{unsrt}

\end{document}